\documentclass[prl,twocolumn,showpacs,preprintnumbers]{revtex4}
\usepackage{graphicx}
\usepackage{dcolumn}

\newcommand{\beq}{\begin{equation}}
\newcommand{\eeq}{\end{equation}}
\newcommand{\beqa}{\begin{eqnarray}}
\newcommand{\eeqa}{\end{eqnarray}}

\begin{document}

\title{Competition between d-wave superconductivity and antiferromagnetism
in the 2D Hubbard model}
\author{M. Capone$^1$ and G. Kotliar$^2$}
\affiliation{$^1$
INFM-SMC and Istituto dei Sistemi Complessi,
Consiglio Nazionale delle Ricerche, Via dei Taurini 19, I-00185, Rome, Italy
and Physics Department, University of Rome ``La Sapienza'',
Piazzale A. Moro 5, I-00185, Rome, Italy}
\affiliation{$^2$ Physics Department and Center for Materials
Theory, Rutgers University, Piscataway NJ USA}

\begin{abstract}
We  study the competition of antiferromagnetism and d-wave
superconductivity at zero-temperature in the two-dimensional
Hubbard model using Cellular Dynamical Mean Field Theory.  The
interplay between the two phases  depends strongly  on the
strength of the  correlation. At strong coupling ($ U \ge 8t$)
the two phases do not mix,  and a first-order transition takes
place as a function of doping between two pure phases.  At
weak-coupling ($U \le 8t$) the two order parameters coexist within
the same solution in a range of  doping and the system smoothly
evolves from the antiferromagnet to the superconductor. When the
transition between the superconducting and the antiferromagetic
phases  is  of the first-order, it is accompanied by a phase
separation.
\end{abstract}

\date{\today}
\maketitle

The competition of d-wave superconductivity (dSC) and
antiferromagnetism (AFM) in the repulsive Hubbard model is a central
problem in the theory of strongly correlated electron systems.
In the cuprate high temperature superconductors dSC emerges 
by doping a parent compound which has AFM long-range order.
Hence, the question of whether the proximity to AFM  is detrimental
or favorable for superconductivity has been debated for nearly
twenty years but no clear consensus has been reached. For a recent
review of these topics see Ref. \cite{demler}.

dSC and AFM can be  clearly identified as the leading
instabilities in weak-coupling functional renormalization group 
of the Hubbard model\cite{metzner}, but this method is unable to describe the competition between 
the two phases at zero temperature. On the other hand, a number of approaches
 predicts that the transition between the AFM and the dSC takes place through an 
intermediate phase where both order parameters are finite.
For example variational wavefunctions of the Gutzwiller type 
give rise to dSC upon doping both in the Hubbard and  $t$-$J$ models\cite{gross}.
However natural extensions of these wavefunctions incorporating the possibility of AFM
\cite{giamarchi} result in a phase diagram where the stable
state is a homogeneous mixture of dSC and AFM.
The slave boson approach to the Resonating Valence Bond (RVB) theory
finds that the dSC state has the lowest energy upon doping in a manifold
of degenerate RVB states at half filling \cite{liu}.
However, when AFM is included into the Hartree-Fock slave-boson decoupling
\cite{doniach}, a mixture of dSC and AFM is stabilized.
Finally the variational cluster perturbation theory approach
 with small cluster sizes find a mixture  of superconductivity and AFM 
away from half filling.\cite{tremblay,arrigoni}

Dynamical Mean-Field Theory (DMFT)\cite{revdmft}, and  its
cluster extensions \cite{clusters} allow us to reexamine the
competition between dSC and AFM from an unbiased perspective in
the sense that all the broken symmetries with order parameters
that fit within a given cluster are treated on the same footing.
Furthermore, by having a  Weiss mean-field containing both
anomalous and normal dynamical components, one expects to avoid
spurious broken symmetries that can appear in
variational treatments with restricted variational freedom.

We consider the two-dimensional Hubbard model
\begin{equation}
H = -t\sum_{\langle i,j\rangle,\sigma} (c^{\dagger}_{i,\sigma} c_{j,\sigma} +
h.c.) + U \sum_i n_{i\uparrow}n_{i\downarrow} -\mu\sum_i n_i,
\label{hamiltonian}
\end{equation}
where $c_{i,\sigma}$ ($c^{\dagger}_{i,\sigma}$) are destruction (creation)
operators for electrons of spin $\sigma$, $n_{i\sigma} = c^{\dagger}_{i\sigma}
c_{i\sigma}$ is the density of
$\sigma$-spin electrons, $t$ is the hopping amplitude, $U$ is the on-site
repulsion and $\mu$ the chemical potential.

In CDMFT we select a cluster of $N_c$ sites, and we map the lattice model 
onto an effective action for the cluster, which 
hybridizes with a bath. A self-consistency equation 
determines the spectral function of the bath, also called dynamical Weiss field.
For practical purposes, it is useful to resort to a Hamiltonian formulation,
where the quantum fluctuations on the cluster are realized by hybridization
with a conduction bath. This leads to a cluster-impurity Hamiltonian of the
form
\begin{eqnarray}
&&\mathbf{H}_{ACI}= H_c + \sum_{k\sigma }\ \varepsilon \,_{k}\
a_{k\sigma }^{+}a_{k\sigma}+\nonumber\\
&&\sum_{k\mu\sigma}\ V_{k\mu\sigma}\ a_{k\sigma }^{+}c_{\mu \sigma} + h.c.+
\sum_{k\mu\sigma}\ V^S_{k\mu}\ a_{k\uparrow }^{\dagger}c_{\mu \downarrow}^{\dagger} + h.c.,
\label{aim}
\end{eqnarray}
where $H_c$ contains the terms of the Hamiltonian which belong to the
cluster, the index $\mu=1,\ldots,N_c$ labels the cluster sites, and
$a_{k\sigma }$ are auxiliary bath degrees of freedom.
Here, since we are interested in superconductivity, there is an
"anomalous" hybridization term which creates and destroys a pair in
which one electron is on the cluster, and the other in the bath.
The cluster-impurity model is still a non-trivial many-body
problem, which we solve using exact diagonalization (ED).
This approach  allows us to obtain zero-temperature results
\cite{caffarel,1d}, and it has been successfully applied to the normal\cite{civelli}
and superconducting state\cite{venky,science_capone} close to Mott insulators.
The use of ED requires a finite Hamiltonian matrix. Hence the sums in Eq.
(\ref{aim}) are limited to a discrete set of values $k_i=1,\ldots N_b$.
This truncation is the only approximation introduced in the ED
approach to CDMFT. In practice, at each DMFT iteration, one determines
the Anderson parameters that better describe the
Weiss field obtained through self-consistency. This requires the 
 minimize a suitably defined distance between the
Weiss field and its discretized counterpart. The details of the implementation
are described in Ref. \cite{1d}.

In this work we always consider a two dimensional $N_c=4=2\times 2$ plaquette
as a minimal cluster where both AFM and dSC are possible, and a bath of $N_b=8$ sites. The same cluster has been studied at finite temperature  within a 
different cluster extension of DMFT in Ref. \cite{sasha}.
Even with this small cluster, the number of parameters to minimize is quite
large.
Since we expect that the main tendencies of the 2D Hubbard model are
AFM and dSC, we found it useful to restrict, as a
{\it preliminary} step to solutions with one or the other broken symmetries.
Therefore we introduced ``constrained" parametrizations with significantly
reduced number of parameters in which only pure AFM
or pure dSC solutions are allowed.
This allows us to determine in a much faster way the regions
of the phase diagrams in which the two pure broken-symmetry phases exist
at $T=0$.
As a second step, we reintroduced the full parametrization, using the
pure solutions defined above as starting points of the iterations, but
adding small perturbations with defined symmetries.
In this way, we have been able to verify the {\it local} stability of the
pure solutions one with respect to the other.
With this second step, we tested the possibility of coexistence between
AFM and dSC.
Finally, we added small perturbations with other symmetries and we also
considered the case with no definite symmetry.

\begin{figure}[tbp]
\begin{center}
\includegraphics[width=7cm,angle=-0] {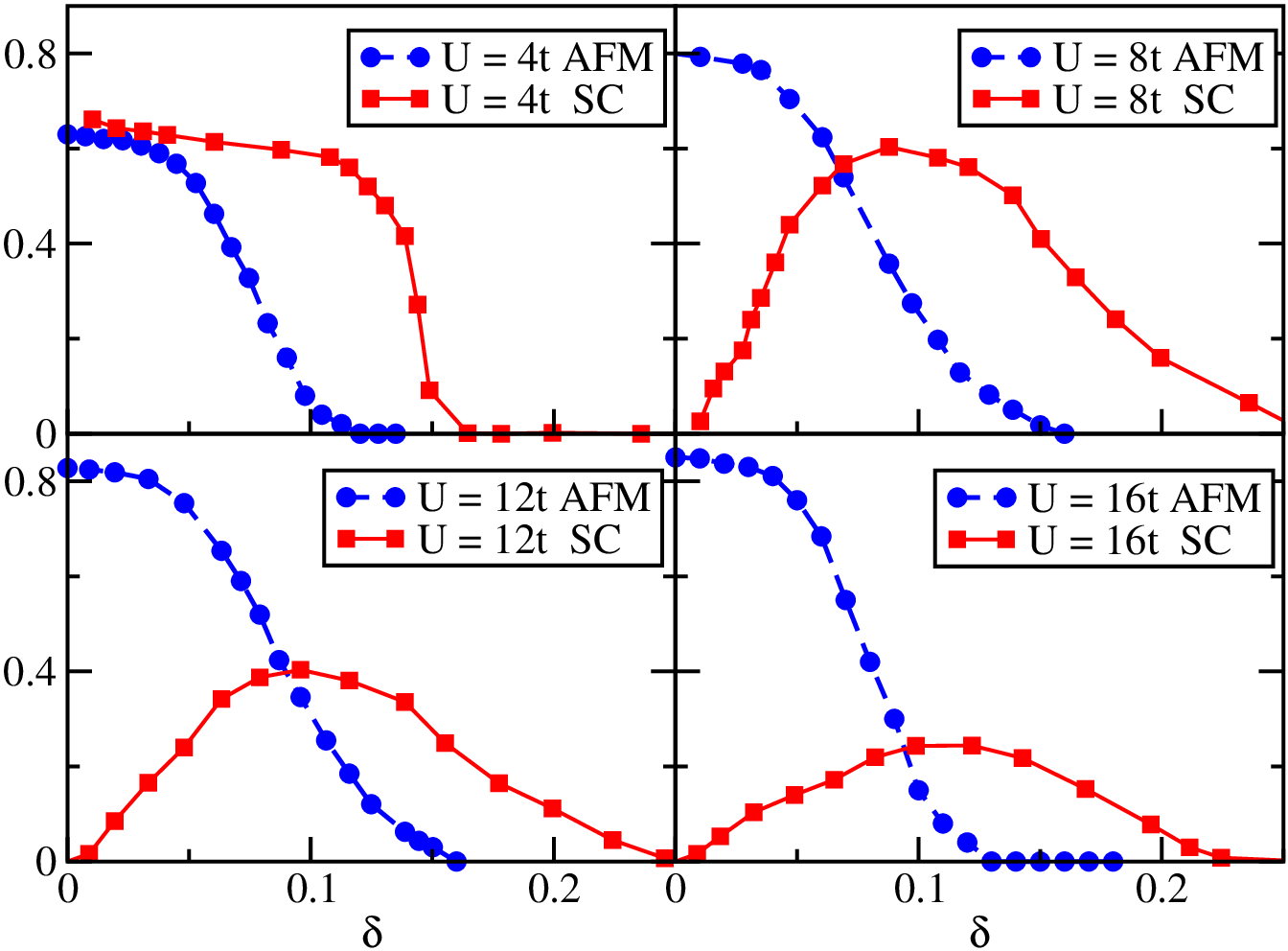}
\caption{(Color online) AFM (blue dashed line with squares) and
dSC (red solid line with circles) order parameters as
a function of doping for four values of the repulsion $U/t = 4,
8, 12$ and $16$. The dSC order parameter is
multiplied by a factor 10 for graphical purposes.} \label{fig1}
\end{center}
\end{figure}

We performed calculations as a function of hole doping $\delta = 1 - n$ 
($n = 1/N_c \sum_{i=1,Nc} \langle n_i\rangle$ being the density per site)
for different values of $U/t$ ranging from weak to strong coupling, and we
applied the protocol defined above.
We first limited ourselves to pure solutions and we identified the regions of
doping in which the pure AFM and dSC solutions exist.
The evolution of the staggered magnetization in the AFM phase $m=\langle \sum_i(-1)^{i}(n_{i\uparrow}i -
 n_{i\downarrow})\rangle$ and of the dSC order parameter in the superconducting phase $\Delta = \langle
\sum_{i} c_{i,\uparrow}c_{i+,\downarrow} -
c_{i,\uparrow}c_{i+y,\downarrow}\rangle$ as a function of doping is shown for $U=4t, 8t, 12t$ and $16t$ 
in Fig. \ref{fig1}.
At half-filling $\delta=0$, the magnetization is an increasing function of
$U/t$ in the whole interaction range. When we dope the system away from
 half filling, $m$ decreases and goes to zero at a relatively large doping
$\delta_{afm} \simeq 0.14-0.16$, which does not depend strongly
on the value of $U$, but has larger values for intermediate
coupling. The behavior of order parameter of
the pure dSC solution is richer. In the  weak-coupling case
$U/t=4$, $\Delta$ evolves in a way similar to $m$, namely it is
maximum at zero doping and monotonically decreases as the hole
concentration grows. The situation changes for large repulsion
values, when the repulsion is large enough to make the system a
Mott insulator even in the absence of any form of magnetic long
range order. In this range $\Delta$  vanishes when we approach the
Mott insulator at zero doping, then it rises   to a maximum and
eventually decreases  for larger doping. Hence  for $U/t= 8, 12$
and $16$,  the  superconducting order parameter  has the
dome-like shape characteristic of  the superconductivity taking
place near a  Mott transition, such as  in the  high temperature
superconductors. The position of the ``optimal doping'' is around
$x=0.1$, and it weakly increases as $U/t$ increases. The maximum
value of the dSC order parameter basically scales with $J =
4t^2/U$ for the large values  $U/t$, as expected in  the
treatments  where the superexchange interaction is the origin of
superconductivity as in the slave boson method. Our pure dSC
solution is similar to that reported earlier in Ref. \cite{venky}.

\begin{figure}[tbp]
\begin{center}
\includegraphics[width=7cm,angle=-0] {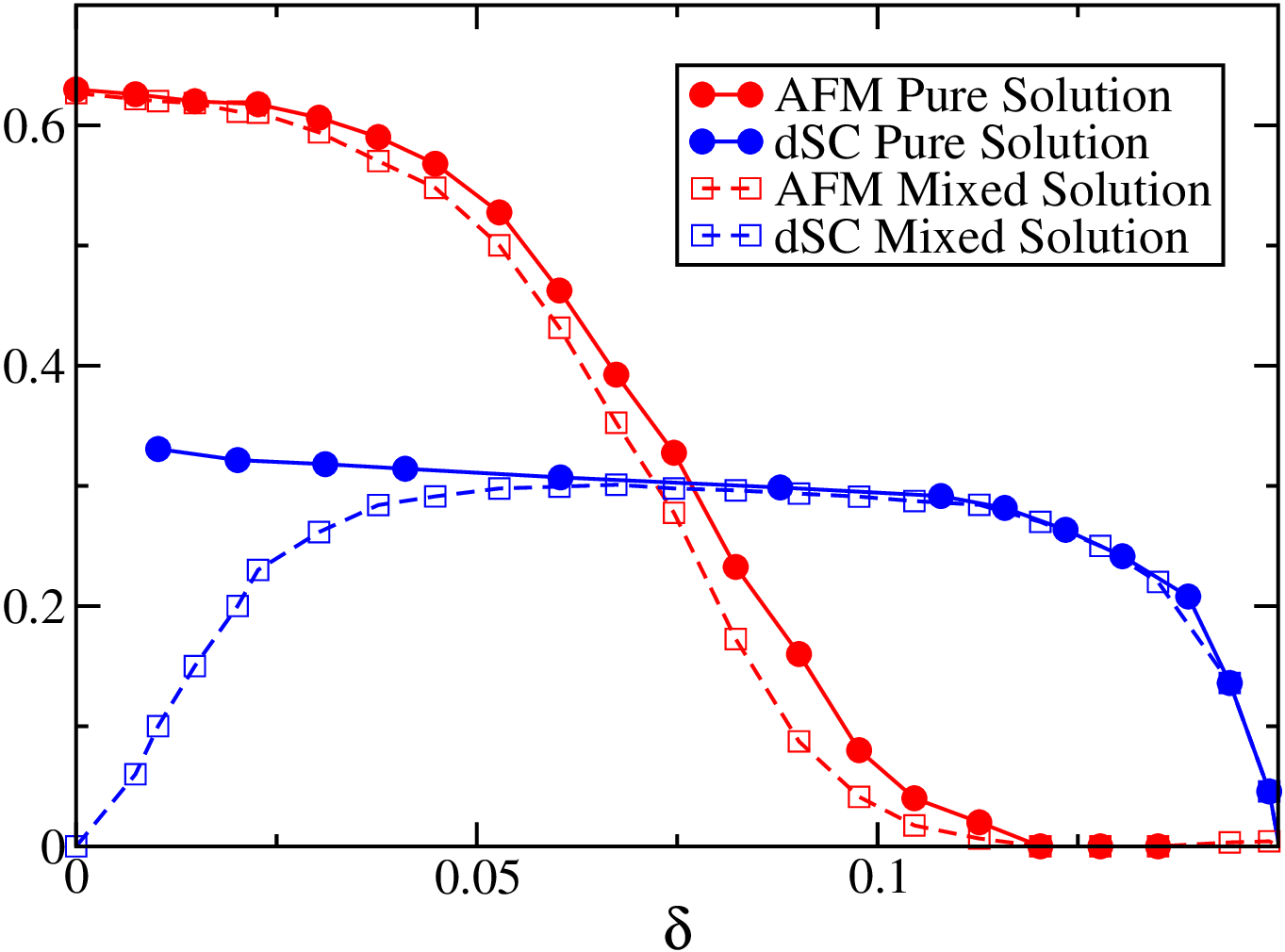}
\caption{(Color online) AFM  and dSC order parameters
as a function of doping for four $U/t = 4$. We compare the values
of the order parameters in the pure solutions (full line and
filled circles) with the values in the mixed solution (dashed
line and open squares). The dSC order parameter was
multiplied by a factor 10 for graphical purposes.} \label{mixing}
\end{center}
\end{figure}

We now turn  to the competition between the two phases. For all
the parameter sets considered above we relaxed the parametrization, allowing for
deviations from  pure AFM or  dSC  order. As mentioned above, we
first considered the local stability of the two solutions.
Therefore, in the regions where two
solutions exist, we started the iterations from one of the pure
solutions, adding a small perturbation in the competing channel.
The outcome of this perturbation strongly depends on the value of
$U/t$. For $U/t=4$, the AFM and dSC solutions, the small
perturbations are stabilized by the CDMFT iterations and a
mixed-phase with both order parameters finite is stabilized. In
Fig. \ref{mixing}, we show the value of the two order parameters
in the mixed state for $U/t=4$, compared with their values in the
starting pure phases. The superconducting component develops for  small
doping, and grows quite rapidly, while the staggered
magnetization is only slightly smaller than the value of the pure
AFM solution. For larger doping the dSC order parameter in the
mixed state  collapses on the pure solution, and the AFM order
parameter becomes slightly smaller. We will see in the following
that the mixed state is not only a solution spontaneously
developed by the iterations, but it also has a lower energy than
the pure ones. A similar behavior has been found in static mean-field 
\cite{bumsoo}.
\begin{figure}
\begin{center}
\includegraphics[width=7cm,angle=-0] {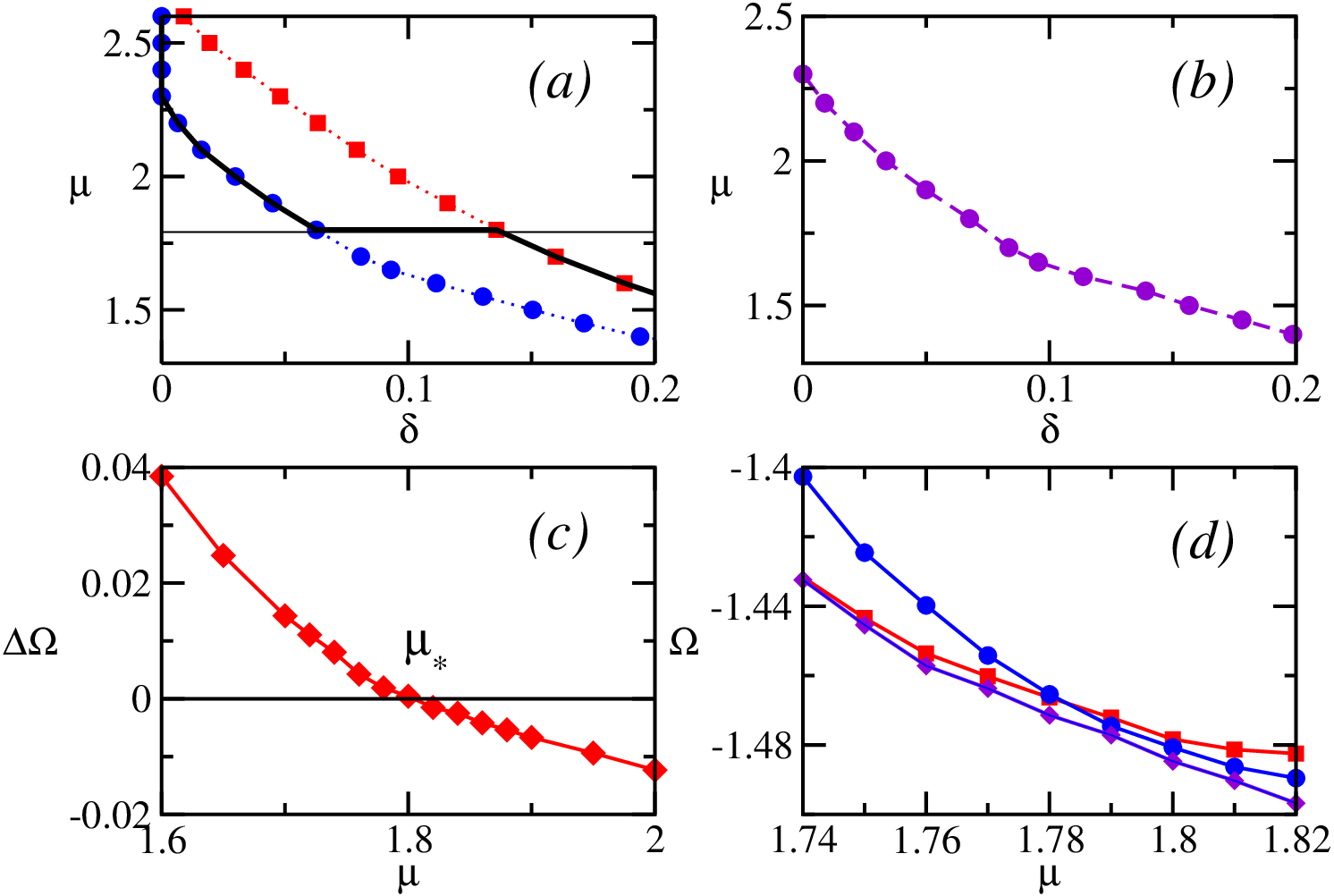}
\caption{(Color online) Chemical Potential as a function of doping for $U=16t$
[panel (a)] and $U=4t$ [panel (b)]. In the former case the red
curve with dots is the AFM solution, the blue one
with squares the dSC, and the black line is the actual solution
as a function of doping. In panel (b) only the stable mixed
solution is shown. Panel (c) shows the difference between the
grandcanonical energies of the AFM and dSC, and Panel (c) compares
the energy of the AFM (red), of the dSC (blue) and of the mixed
solution (violet).} \label{fig3}
\end{center}
\end{figure}

For large $U/t = 12, 16$, the two  solutions  (pure dSC and pure
AFM) are  found to be stable against the  perturbations described
above: a small dSC(AFM) perturbation of the AFM(dSC) state
disappears as the iterative procedure goes on, signaling that the
two states  are in direct competition which each other and cannot
be connected. For $U/t = 8$ the situation is intermediate: A
small mixture between the two solutions takes place, but the
magnitude of the ``minority'' order parameter is found to be
really small, basically of the same order of the truncation error
of the ED calculation. For this reason we can not judge whether
this value of $U$ lies in the weak-coupling or in the
strong-coupling regime, but we can surmise it is close to the
boundary between the two regimes.

The third step of our approach      perturbs the previously
obtained solutions with Weiss fields  with other symmetries (e.g.,
a $d+is$ superconductivity) or with  perturbations with no
definite symmetries. Regardless the value of $U/t$, we find that
these perturbations all vanish through the iterative procedure.
Thus we conclude that we only have pure AFM, pure dSC and mixed
AFM+dSC  as locally stable solutions of the CDMFT equations on a
two by two plaquette. To establish which  solution is the globally stable one  we
compute their  grandcanonical
potential at zero temperature  $\Omega = \langle H - \mu N\rangle = 
\langle H_{kin} \rangle + \langle H_{int} \rangle -\mu\langle N\rangle $.
The interaction term is given by the expectation value of the
double occupancy on the cluster sites, while the kinetic energy
requires the knowledge of the lattice Green's function
$G(k,\omega)$, as $E_{kin} = \sum_k \epsilon_k G(k,\omega)$,
where $\epsilon_k$ is the non-interacting dispersion. 
Different schemes have been proposed to extract lattice properties
from cluster ones. Here we use the
approach of Ref. \onlinecite{cdmft}, where the lattice
self-energy is obtained by periodizing the cluster
self-energy, but  we  have also checked that alternative methods
\cite{tudor,tremblay,venky} do
not qualitatively affect the qualitative phase boundaries and the
nature of the transitions, and that the quantitative
differences are not large.

For $U/t=4$ the mixed phase has lower
$\Omega$  than the two pure phases, and it is thermodynamically 
stable since the 
chemical potential is a monotonic function of the doping (panels (b-d) of Fig. \ref{fig3}).
For $U/t=12$ and $16$, the comparison of the energy determines the range of
absolute stability of the two mutually exclusive solutions, as shown in panel (c) of Fig. \ref{fig3},
where the difference $\Delta\Omega = \Omega_{dSC} - \Omega_{AFM}$ is plotted for $U/t = 16$.
$\Delta\Omega$ becomes zero at  $\mu = \mu^*$, where the system 
jumps from one phase to the other through a  first-order transition.
the curve of $\Omega(\mu)$ has the wrong curvature (corresponding to negative compressibility)
in an interval. The system is therefore no longer stable in a
uniform phase, and it phase separates. This is clearly seen in the plot of the chemical potential
as a function of doping, where at  $\mu^*$ the system turns from AFM to dSC, leaving a window of
forbidden dopings.

In conclusion, we have studied the competition of AFM and dSC within  the
two-dimensional Hubbard model at zero  temperatures using the
CDMFT, an unbiased approach which treats both forms of
order on the same footing.
 Using an ED solver, we have been able to
span a wide range of correlation values, from $U/t=4$ to
$U/t=16$, and followed the evolution of the two phases.
At weak-coupling the AFM and dSC coexist in a single
phase with two order parameters in agreement  with the results of
a weak coupling analysis \cite{sachdev}. In the pure solutions
with a single order parameter, $\Delta$ and $m$ are
maximum at half-filling and decrease as a function of doping.
Allowing for a simultaneous ordering suppresses the dSC
order parameter in favor of the AFM one.
For large $U$ we  find that there is no
mixture and AFM and dSC exist as pure
phases. A first-order transition as a function of doping 
occurs between the two phases, accompanied by a phase
separation. Hence AFM and dSC exclude each other for large $U$. 
It is therefore possibile to follow the metastable 
pure d-wave state at small doping to study the
approach towards  the Mott insulator. This will be explored in a
forthcoming publication \cite{science}. While many properties  of
the  pure superconducing phase  are correctly described by simpler
methods such as  Gutzwiller variational wavefunctions or slave
boson techniques, these   methods are not able to describe
the stability against admixture of magnetism which requires an
accurate description of both ordered phases. In fact, the abrupt
first order phase transition between dSC and AFM can only been obtained with fairly
sophisticated trial wavefunctions \cite{shih} or through  the bond
operator method, which  captures some aspects of the CDMFT with static
expectation values \cite{sachdev}.
It is interesting to notice, that a  condition for the Hubbard
model to have  SO(5) symmetry requires  to be in the tricritical
region where the  competition between AF and dSC  switches to an
homogeneous mixture of the two order parameters \cite{demler}. In
our calculations this occurs in the interesting intermediate
coupling regime  of $ U \approx 8 t$.

We emphasize here that  CDMFT   is designed to   provide the best
possible description of the dynamics associated with short-range
correlations. For this reason we restricted our study to
homogeneous states. CDMFT   indicates that the pure
Hubbard model should display phase separation in agreement with
many other methods \cite{kivelson}. These tendency can be
enhanced or eliminated by including more realistic features
such as longer range interactions or hoppings, that may lead to 
more  general real-space patterns such as stripes.  This
study clearly requires large clusters, or a long wavelenght
Landau-Ginzburg approach,  which could exploit the CDMFT $2\times 2$ results as
an input for determining the effective parameters.
On the other hand, the different nature of the interplay of
AFM and dSC in weak and strong
coupling pointed out here is expected to be a very robust
feature   captured by our treatment of the Hubbard model.
Furthermore,  we expect that the basic energetics will have a
weaker dependence on  additional interactions such as
the coupling to phonons that might be required for a realistic
modeling of a specific correlated material.

M. C. acknowledges the warm hospitality of Rutgers
University. Useful discussions with M. Civelli and A.-M. Tremblay
are gratefully acknowledged.  This work was supported by the NSF
under grant DMR 0528969, the Italian Miur Cofin 2005 and CNR-INFM.


\end{document}